\documentclass[%
 aip,
 amsmath,amssymb,
 preprint,%
]{revtex4-1}

\draft 
\usepackage{graphicx}
\usepackage{dcolumn}
\usepackage{bm}
\usepackage[utf8]{inputenc}
\usepackage[T1]{fontenc}
\usepackage{mathptmx}
\usepackage{etoolbox}

\usepackage{amsbsy}
\usepackage{amssymb}
\usepackage{amsmath}
\usepackage{mathtools}
\usepackage{color}
\usepackage{multirow}
\usepackage{rotating}
\usepackage{url}
\usepackage{indentfirst}
\usepackage{xcolor}
\usepackage{wrapfig}
\usepackage{subcaption}

\begin{document}

\title[Deep reinforcement learning for flow control: perspectives and future directions]{Recent advances in applying deep reinforcement learning for flow control:\\ perspectives and future directions}
\author{C. Vignon}
 \affiliation{ 
FLOW, Engineering Mechanics, KTH Royal Institute of Technology, SE-100 44 Stockholm, Sweden
}%
 \affiliation{ 
Mines de Paris - Université PSL, 75005 Paris, France 
}%
\author{J. Rabault}%
\affiliation{ 
IT Department, Norwegian Meteorological Institute, Postboks 43, 0313 Oslo, Norway
}%

\author{R. Vinuesa}
 \affiliation{ 
FLOW, Engineering Mechanics, KTH Royal Institute of Technology, SE-100 44 Stockholm, Sweden
}%

\date{\today}

\begin{abstract}
Deep reinforcement learning (DRL) has been applied to a variety of problems during the past decade, and has provided effective control strategies in high-dimensional and non-linear situations that are challenging to traditional methods. Flourishing applications now spread out into the field of fluid dynamics, and specifically of active flow control (AFC). In the community of AFC, the encouraging results obtained in two-dimensional and chaotic conditions have raised interest to study increasingly complex flows. In this review, we first provide a general overview of the reinforcement-learning (RL) and DRL frameworks, as well as their recent advances. We then focus on the application of DRL to AFC, highlighting the current limitations of the DRL algorithms in this field, and suggesting some of the potential upcoming milestones to reach, as well as open questions that are likely to attract the attention of the fluid-mechanics community.
\end{abstract}

\maketitle

\section{\label{INTRO}Introduction}

Deep reinforcement learning (DRL) is a combination of deep learning (DL) and reinforcement learning (RL). Its ability to solve high-dimensional, non-linear complex problems has been proven over the past decade in multiple domains such as, non exhaustively, robotics~\cite{2013Kober, Gu2017,Andrychowicz}, language processing~\cite{Li2016}, board games (e.g., the game of Go~\cite{Silver2017MasteringGo, Silver2016}), Poker~\cite{BrownSandholm2019}, video games~\cite{2013Mnih, 2015Mnih, Szita2012} and image analysis~\cite{2012Krizhevsky}. The acronym DRL is used when a deep neural network (DNN) takes place in the RL process. DNNs are well-known for their ability to approximate non-linear functions on high-dimensional spaces~\cite{Hornik1989MultilayerFN}, and have been introduced in RL methods for that particular purpose. Thanks to these artificial neural networks (ANNs), the algorithms of the 1990s solving low-dimensional problems~\cite{Tesauro1995TemporalDL, Bagnell2001_appliRL, Ng2004_appliRL, PetersSchaal2006_appliRL, PetersSchaal2008_appliRL, Diuk2008_appliRL, Riedmiller2009} have been enhanced and now reach super-human performances in games and can solve a broad range of non-linear, high dimensional optimization tasks in various domains, as cited above.

Reinforcement-learning methods~\cite{SuttBarto2018} rely on the interaction between an agent and an environment. The agent receives observations from the environment and then decides of an action according to those. Its aim is to maximize a reward that characterizes the 'goodness' of the state of the environment. If the agent interacts with a model of environment instead of the environment itself, the RL algorithm is said 'model-based', otherwise if the agent interacts directly with the environment, the RL algorithm is 'model-free'. In that case, the algorithm only uses the partial observations of the environment as inputs, and returns the actions as an output. Such algorithms are thus considered as 'black-box' methods, because an analytical description of the system is not needed to control it. 
\\

Problems including chaotic (turbulent) flows are usually non-linear and high-dimensional, due to the governing Navier--Stokes equations and the continuous environment space, i.e. the flow. Efficient 'black-box' RL methods can therefore be useful and fruitful in that field, as traditional analysis methods based on local linearization usually struggle in this context. In Ref.~\onlinecite{BruntonNoack2015}, the authors review the main methods of closed-loop RL adapted to the specific case of fluid dynamics and their inherent challenges, i.e. high-dimensionality of the environment and/or action space(s), and complex non-linearity of the governing equations. The application of RL methods and derivatives in fluid mechanics has known a flourishing growth for a few years. One may refer to recent advances with late reviews on DRL~\cite{2021Garnier, 2022Vinuesa}, and more general machine-learning (ML) methods applied to fluid dynamics~\cite{2022Pino}. The use of RL in fluid mechanics has various purposes, such as (non-exhaustively): subgridscale~\cite{petros_natmi,KURZ2023109094}, wall~\cite{Bae2022} and turbulence~\cite{2019Beck, Beck2019_closure_turbModel, Beck_2021_TurbulenceModelling} modelling, RL-augmented computational fluid dynamics (CFD) solvers~\cite{KURZ2022_RLagumented_CFD, Kurz2022_Relexi}, the shape optimization of airfoils~\cite{2008Lampton, Lampton2010, 2021Ghraieb}, the active control of the separation in a flow surrounding an aircraft~\cite{batikh:hal-01820331}, the reduction of the drag past cars~\cite{noack2021_DragRed_carsTrucks} and bluff bodies in both laminar~\cite{2019Rabault-al} and chaotic regimes~\cite{2020Tang_RobustAFC, 2021Ren_AFCturb, 2022Varela}, the study of the wake behind a cylinder~\cite{2021LiZhang} or the reduction of the skin-friction drag in a turbulent channel~\cite{2022Hasegawa,Guastoni2023}. Within the non-RL machine-learning methods, genetic programming~\cite{1992Koza} (GP) has shown comparative results in multiple applications~\cite{gautier2015, 2017Li_GP_dragRed, FanNoack2019, Ren2019, Hao2020}. For the sake of brevity, as an entire review could be dedicated to those, GP algorithms will be briefly introduced in the present review solely to compare their main properties with the DRL ones.

In fluid mechanics, two fields are therefore mainly studied for the application of RL: shape optimization and flow control. Within the former, a body is immersed in a flow, and the RL algorithm modifies its shape in order to optimize some characteristic due to the dynamic of the fluid (e.g., the drag and/or the lift -- see for instance Ref.~\onlinecite{Ghraieb2022_shapeOpt}). A classical example is the reduction of the drag coefficient behind a body in a laminar or turbulent flow: in Refs.~\onlinecite{2008Lampton,Lampton2010} this idea is applied to the shape optimization of airfoils (more recent studies on that subject are given in Refs.~\onlinecite{2021Ghraieb,2020Viquerat}, using DRL).
Flow control aims at actively influencing the behavior of a flow, such as controlling separation, turbulence or heat transfers, in order to reach a more desirable state. It is a subject of great interest from both societal and economical points of view~\cite{2021Ghraieb}. Flow control strategies can be passive or active~\cite{GadElHak96}.  Contrarily to the passive solutions (e.g. riblets and vortex generators~\cite{COLLIS2004}), the active methods need a supply in energy to perform the control sequence. As they could achieve unrivalled performances, DRL algorithms applied to active flow control problems are largely studied for their promising engineering prospects (see, e.g., Refs.~\onlinecite{BruntonNoack2015,2022Vinuesa}).
\\

This review is aimed at presenting the current challenges in the field of active flow control (AFC), particularly when the control is based on DRL methods. To this end, a general overview of the main reinforcement-learning processes is first given in section \ref{RLpresentation}, and then follows a presentation of the recent advances in deep reinforcement learning (section \ref{sec:DRL}). Eventually, applications of machine learning -- and specifically DRL -- in AFC are developed in section \ref{sec:DRL_and_flow_control}. Attention will be focused on the challenges inherent to the field of fluid dynamics, and some possible upcoming milestones for the coupling of DRL and AFC will be suggested accordingly. 

\section{Reinforcement Learning} \label{RLpresentation}

\subsection{Introduction}

Reinforcement learning (RL) is a branch of data-driven methods, considered as a subsection of machine learning, which is itself a subset of artificial intelligence. The first study developing the modern form of reinforcement learning was provided by \citet{sutton1988learning} at the end of the 1980s. In his time, \citet{sutton1988learning} merged two distinct approaches: the learning by trials and errors, notably studied for its natural applications in the learning process of animals, and optimal control. By combining the two, the aim of modern RL is to optimize the decisions taken by an agent in a surrounding environment during a sequence of actions. In other words, the idea is to train and improve a control law (the agent) able to optimize a target system (the environment) by taking sequences of actions.

Reinforcement learning has already been applied to many fields with successful results. During the past decade, RL algorithms have been enhanced thanks to neural networks, developing DRL methods on the basis of RL. One may refer to Ref.~\onlinecite{2013Kober} for RL and Ref.~\onlinecite{Andrychowicz} for DRL applied to robotics, to Refs.~\onlinecite{2013Mnih,2015Mnih} for DRL applied to Atari games with super-human performances, to Ref.~\onlinecite{Szita2012} for video games and Ref.~\onlinecite{Silver2016} for Go game, to name a few.
\\

In reinforcement learning, an agent interacts with an environment in a three-steps sequence : 
\begin{enumerate}
    \item The environment gives to the agent a partial observation $o$ of its state $s$. In a continuous environment space, the observation is necessarily incomplete. Thus, the observation $o$ may not provide enough information to describe $s$. In order to overcome this issue, the state $s_t$ obtained at time $t$ is often described by the succession of actions and observations that led to it : $s_t~=~(s_0,a_1,o_1,a_2,...,o_{t-1},a_t)$, or similarly: $s_t~=~(s_0,a_1,s_1,a_2,...,s_{t-1},a_t)$.
    \item According to the observation $o$ of state $s$, the agent takes an action $a$, following a certain policy $\pi$: $a = \pi(s)$. This action impacts the environment and modifies its state. As a consequence, its new state is now denoted $s'$ (\textit{a priori} different from $s$).
    \item Eventually, a reward $r$ is given to the agent, characterizing the 'goodness' of the action taken according to the old state $s$ and the new one $s'$.
\end{enumerate}

\begin{figure}
\includegraphics[width = 0.5\textwidth]{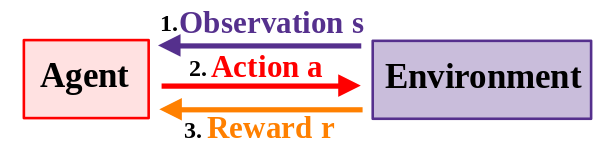}
\caption{\label{fig:schemeRL}Interactions between an agent and an environment in a generic reinforcement-learning loop.}
\end{figure}

Through these three interactions, which take place at each time-step of a global process, the RL algorithm gathers information and progressively enhances the reward. To this end, the reward defines the feature that the user wants to improve. Instead of the instantaneous reward, introduced as $r$ above (or more precisely $r_t$ at time $t$), RL algorithms usually optimize a cumulative reward, denoted by $R$ and defined as:
\begin{equation}\label{eq:FirstCumuReward}
    R = \sum \limits_{{t=0}}^T \gamma^t r_t \text{   },
\end{equation}
where $\gamma$ is the discount factor (usually, $\gamma \gtrsim 0.99$), and $T$ is the duration of an episode. The sequence of the algorithm is as follows:
\begin{itemize}
    \item First, at $t=0$, the state of the environment is (randomly) initialized (state $s_0$).
    \item At each time-step $t$, the three steps process described above (see Fig.\ref{fig:schemeRL}) is applied, returning a reward $r_{t}$ after having changed the state of the environment from $s_{t}$ to $s_{t+1}$, by enforcing the action $a_{t}$ according to the policy $\pi$.
    \item When $t$ reaches $T$, the process stops. The sequence of actions running during a time $T$ is called an episode. At the end of an episode, the cumulative reward $R$ is computed and the policy modified with the long-term objective of optimizing $R$. During an episode, $\lfloor \frac{T}{t} \rfloor$ actions are taken. In the cumulative reward, actions with a low instantaneous reward can be compensated by others, no distinction is made apart from the discounted factor $\gamma^t$. This factor slightly favours short-term rather than long-term strategies, while enabling $R$ to converge when episodes with numerous actions are considered. A full sequence $(s_0, a_1, s_1, a_2, ..., s_T)$ is called a trajectory, and further represented by $\tau$. 
\end{itemize}

The cumulative reward is commonly not computed over the whole trajectory, but from a certain time-step $t_0$. Equation (\ref{eq:FirstCumuReward}) is then replaced by:
\begin{equation}\label{eq:secondCumuReward}
    R = \sum \limits_{{t=t_0}}^T \gamma^{t-t_0} r_t~.
\end{equation}

Depending on the representation chosen for the environment, RL algorithms are ordinarily separated into two broad categories. If the agent interacts with an artificial modelling of environment, the algorithm is considered to be model-based, otherwise the algorithm is model-free. In section \ref{sec:Model_based}, examples of model-based methods are presented, but since the majority of DRL algorithms are model-free, attention will mainly be given to the model-free category in this review. Additionally, the reader can refer to the works of \citet{jaderberg2017} and \citet{2018HaSchmidhuber}, or to the recent review of \citet{RabaultKuhnle2022_book}, for examples of model-free algorithms incorporating a model-learning process. By trying to build a representation of the environment during the learning process and benefit from it, these initially-model-free methods are at the cross-section of model-free and model-based algorithms.
\\

In the following subsections we will only discuss the main concepts and features of RL algorithms, without a focus on DRL. How NNs can be used together with RL algorithms to develop DRL will be discussed in section~\ref{sec:DRL}.

\subsection{Model-based methods}\label{sec:Model_based}
Model-based methods require an artificial representation of the environment. Historically, linear models first spread across the field of flow control, as they were relatively simple and proved to be efficient on multiple problems, e.g the stabilization of convectively unstable flows~\cite{Barbagallo2012,dahan_morgans_lardeau_2012, Gautier2014FeedforwardCO}. But flows are also commonly characterized by their strongly nonlinear behaviors. Although linear models are justifiable in turbulent situations where the flow behaves like a linear amplifier~\cite{brackston2016}, their application to highly turbulent regimes is ambiguous and has long been debated in the community (see, e.g., Ref.~\onlinecite{KimBewley2007} and the references therein). To that purpose, \citet{brackston2016} developed a (nonlinear) stochastic model, inspired from the work of \citet{2015Rigas} and governed by the nonlinear Langevin equation. A feedback controller coupled to this model successfully stabilized the wake behind a bluff body immersed in a three dimensional turbulent flow. The drag was efficiently reduced by suppressing the large-scale structures. In their review, \citet{RowleyDawson2017} analyze the robustness of reduced order models when applied to linear and nonlinear problems. These model-based methods aim at extracting the main dynamic processes that structure the flow, in order to simplify its modelling. The reader may also refer to the works of \citet{Queipo2005} and \citet{Koziel2016} for examples of model-based applications in aerodynamics.
\\

The computational gain obtained with model-based methods results from the simplification of the environment. This gain is especially relevant in the situation of flow control, but necessarily leads to sub-optimal solutions when it comes to high-Reynolds-numbers regimes, as the models only partially approximate the turbulences. Hence, models can still be configured for these ranges of regime, but one is then confronted to a mediation between sub-optimal complex model-based controllers that rely on necessarily approximate models of the flow, and computationally-heavy 'black-box' model-free controllers that directly interact with a Navier-Stokes solver.
Recent model-based methods, such as the Probabilistic Ensembles with Trajectory Sampling (PETS)~\cite{2018Chua}, try to compete with the efficient modern model-free deep RL algorithms and thus go along the direction of developing both model-based and model-free methods.

\subsection{Model-free methods}\label{sec:Model_free}
Model-free methods are increasingly popular in the community because of how relatively simple they are to apply. Indeed, the agent directly interacts with the environment: no assumption has to be made on the environment modelling. Thanks to regular partial observations of the environment state, the agent finds a control sequence $(a_0, ..., a_t, ..., a_T)$ that maximizes the cumulative reward.

The environment is usually considered as stochastic (i.e. when the action $a$ is applied on the state $s$, the new state cannot be predicted and deduced exactly) and is commonly modelled by a Markov decision process (MDP)~\cite{Bellman1957, howard1960, SuttBarto2018}, even though it could be represented by other stochastic processes~\cite{2019Rabault-al}, or considered to be deterministic~\cite{Silver2014DeterministicPG}.

\subsubsection{Markov Decision Process}

Markov decision processes (MDPs)~\cite{Bellman1957, howard1960} are at the core of modern reinforcement learning~\cite{SuttBarto2018}. A MDP is a tuple $(S,A, \mathcal{P}_a, R)$ with $S$ and $A$ the sets of (respectively) all the possible states and actions, and $R$ the set of all the possible rewards. One can define $\mathcal{P}_a(s,s')$ as the probability that taking action $a$ in state $s$ will lead to state $s'$. Reusing the standard nomenclature proposed in Ref.~\onlinecite{2021Garnier}, let us define the operator $\mathbb{P}$ as:
\begin{eqnarray}
        \mathbb{P}: S\times A \times S \longrightarrow [0;1] \text{ };\nonumber\\ 
        (s,a,s') \mapsto \mathbb{P}(s'|s,a) = \mathcal{P}_a(s,s')~.
\end{eqnarray}

Hence, $\mathbb{P}$ represents the probability of obtaining state $s'$ after having taken action $a$ in state $s$. The policy is defined by:
\begin{eqnarray}
\pi_{\theta}:S \longrightarrow A \text{ };\nonumber\\
s \mapsto \pi_{\theta}(s)~.
\end{eqnarray}

Note that $\pi_{\theta}(s)$ returns the actions probability distribution when the initial state is $s$. The action is then chosen by the agent according to $\pi_{\theta}(s)$, and $\theta$ represents the set of parameters that characterizes $\pi$. The aim is to find the policy $\pi$ that maximizes the cumulative reward $R$. Modifications of the policy are made by choosing different sets of parameters $\theta$. Furthermore, $\mathbb{P}$ and $\pi_{\theta}$ completely describe the fundamental process of the RL algorithm: given a state $s$, the agent takes the action $a$ according to $\pi_{\theta}(s)$, and then $\mathbb{P}$ returns the probability of obtaining a new state $s'$.

\subsubsection{State value and state-action value functions}

In addition, three functions are usually defined in order to describe and classify the RL algorithms.
The state-action value function, also called the \textit{$Q$-function}, is defined by:
\begin{equation}
    Q^{\pi}(s,a)= \mathbb{E}_{\tau \sim \pi}\left[R(\tau)|s_0 = s,a_0=a\right]\text{ },
\end{equation}
and the state value function is defined by:
\begin{equation}
    V^{\pi}(s)= \mathbb{E}_{\tau \sim \pi}\left[R(\tau)|s_0 = s\right]\text{ }.
\end{equation}

Hence, $Q^{\pi}(s,a)$ corresponds to the expected cumulative reward when the action $a$ is taken in the initial state $s$ and then follows the trajectory $\tau$ according to the policy $\pi$, whereas $V^{\pi}(s)$ corresponds to the expected cumulative reward when the initial state is $s$ and then follows the trajectory $\tau$ according to the policy $\pi$. Thus:
\begin{equation}
    V^{\pi}(s)= \mathbb{E}_{\tau \sim \pi} [Q^{\pi}(s,\pi(s))]~.
\end{equation}

Eventually, the advantage function $A^{\pi}$ is defined by:
\begin{equation}\label{eq:advFunct}
    A^{\pi} = Q^{\pi} - V^{\pi}~.
\end{equation}

$A^{\pi}$ characterizes the advantage (in terms of cumulative reward) of taking action $a$ in state $s$ rather than taking any other possible action.

Among the model-free RL algorithms, two main methods are both developed for their promising performances: value-based~\cite{Watkins89} and policy-based methods~\cite{Williams1992}. During the RL process, the agent takes actions according to the state of the environment by following a policy. With policy-based methods, the aim is to directly optimize the policy $\pi$ (see section \ref{sec:PB_methods}). Differently, value-based methods rely on learning the Q-function, and then deriving the optimal policy from it (see section \ref{sec:VBmethods_RL}).

\subsubsection{Closed-loop and open-loop methods}

Additionally to the value/policy criterion, RL methods are also commonly distinguished according to the frequency of the interactions between the agent and the environment. In closed-loop control, the agent obtains a regular feedback of the current state of the environment thanks to sensors. The RL algorithm uses these measurements to adapt its policy during the episode, in order to tend to a more desired state (see Ref.~\onlinecite{BruntonNoack2015} for a review on closed-loop turbulence control). On the other hand, open-loop methods assume that the decision policy does not rely on the state of the environment, or at least that a steady or periodic actuation can modify the principal dynamical processes of the environment~\cite{Meliga2010}. One may refer to the recent works of \citet{Shahrabi2019} and \citet{2021Ghraieb} for the use of open-loop methods in flow control. By not considering the surrounding environment, these methods are generally less efficient than the closed-loop solutions, but easier to implement. The upcoming sections will mainly focus on closed-loop algorithms, as they are the most promising and challenging methods for the future.

\subsubsection{On-policy and off-policy online and offline methods}

Another distinction among RL algorithms is made between on-policy and off-policy methods. Within the former, the agent solely learns about the policy it enforces to the environment, contrarily to the latter where the agent can learn from other policies. For instance, \citet{Degris2012b} execute a second (behavior) policy aimed at choosing the trajectories while the first one decides the actions. Another example of off-policy method is the Q-learning process~\cite{WatkinsDayan1992} mentioned in the next section. Furthermore, with on-policy methods, the data collected before the last update of the (single) policy cannot be used for forthcoming updates, as these data had been generated by a policy that differs from the last version. Therefore, off-policy settings are of great interest, as they enable the algorithm to reuse previous experiences that may be stored from either earlier in the learning process or completely separate learning runs, and to learn multiple tasks in parallel thanks to the different policies.

Offline RL algorithms learn and update their policy(-ies) only at the end of an episode, or after N episodes, while online algorithms make updates 'in real time'. Online strategies present the advantage to learn from the current state of the environment instead of past situations, but require demanding computational properties~\cite{Sutton1993OnlineLW}, not always available nor achievable. The reader may refer to the work of \citet{Degris2012a} and the references therein for additional information on the differences between offline and online methods.

\subsubsection{Value-based methods}\label{sec:VBmethods_RL}

Value-based methods rely on the Q-function. 
The optimal Q-function can be defined as $Q^*~=~\text{max}_{\pi}Q^{\pi}$. In reinforcement learning, the objective of value-based methods is to have a Q-table filled with the values of $Q^*(s,a)$ for any $(s,a)$ in $S \times A$. In other words, the aim is to be able to know the best policy to apply when starting from any tuple $(s,a)$ in $S \times A$, in order to maximize the expected cumulative reward. The Q-table can be thought of as an array or a transition reward matrix in discrete states case, or can be approximated by a NN or another function approximator in general. For simplicity, in this section Q is considered as an array that can be filled with values, while an ANN trained by gradient descent will be considered in deep Q-learning (see section \ref{sec:DQL}).

It is not possible to directly compute $Q^*$, contrarily to $Q^{\pi}$. Thus, the Q-table is first filled with values of $Q^{\pi}$, estimated by exploring the space $S \times A$. Additionally, $Q^*$ respects the Bellman optimally condition~\cite{Bellman1962}:
 \begin{eqnarray}\label{eq:bellmanOptimallyCond}
     Q^* (s,a) = r(s,a) + \sum \limits_{{s' \in S}} \mathbb{P}(s'|s,a)\gamma \text{max}_{a'}Q^*(s',a') \nonumber\\
     = r(s,a) + \mathbb{E}_{s' \in S}\left[\gamma \text{max}_{a'}Q^*(s',a') \right]~.
 \end{eqnarray}

Here, $r(s,a)$ represents the instantaneous reward returned when action $a$ is taken in state $s$. This relationship on $Q^*$ means that when the action $a$ is first applied to the initial state $s$, leading to the state $s'$ (as it is a stochastic process, $s'$ is not known precisely, hence the expectation on $s'$), if the optimal Q-function is known at the next time-step, then the best policy is to choose the next action $a'$ that maximizes $Q^*(s',a')$.
 
To update the Q-table, Eq.~(\ref{eq:bellmanOptimallyCond}) inspires the recursive update relationship given below:
\begin{equation}\label{eq:RecursiveBellman}
    Q^{\pi}(s,a) \longleftarrow \mathbb{E}_{s' \in S}\left[r(s,a) + \gamma \text{max}_{a' \in A}Q^{\pi}(s', a')\right]
\end{equation}

 It has been proved that updates of $Q^{\pi}$ by following (\ref{eq:RecursiveBellman}) make $Q^{\pi}$ converge to $Q^*$~\cite{Bellman1962}.

Eventually, when the Q-table is filled with the estimated values of $Q^*$ (thanks to the recursive process given by Eq.~(\ref{eq:RecursiveBellman})), and given a state $s$, one can then optimize the action that the agent will take, by choosing the action $a^*$ according to the 'greedy' policy:
\begin{equation}
    a^* = \text{argmax}_{a \in A} Q^*(a, s)~.
\end{equation}

These value-based methods are usually called Q-learning~\cite{WatkinsDayan1992}. One can refer to the original work of \citet{Watkins89}, which was already an evolution of difference learning~\cite{sutton1988learning}.

\subsubsection{Policy-based methods}\label{sec:PB_methods}

The second main category of RL methods directly relies on an estimation and an optimization of the policy function $\pi_{\theta}$, and does not consider an optimization of the $Q$-function. These policy-based methods have been developed since -- at least -- the work of \citet{Williams1992}. The objective function $J$ is introduced in order to evaluate the quality of the policy. As the long-term aim still remains to maximize the expected cumulative reward, $J$ is commonly defined by:
\begin{equation}
    J(\theta) = \mathbb{E}_{\tau \sim \pi_{\theta}}\left[R(\tau) \right]~.
\end{equation}

$J$ can vary thanks to modifications of $\theta$, the set of parameters characterizing the policy. Thus, the objective is to find:
\begin{equation}
    \theta^* = \text{argmax}_{\theta}J(\theta)~.
\end{equation}

To that purpose, the gradient of $J$ is introduced, so that $\theta$ can be recursively modified by following the direction of this gradient: 
\begin{equation}\label{eq:updateThetaRec}
    \theta \longrightarrow \theta + \lambda \nabla_{\theta}J(\theta)~,
\end{equation}
with $\lambda$ being a positive parameter. Two distinct options are discussed for the evaluation of $\nabla_{\theta}J$, whether the policy is considered to be stochastic or deterministic, i.e. whether $\pi(s)$ returns a probability distribution of actions $a$ or an exact and unique value $a$. In the former case, the gradient is computed according to the relationship~\cite{2021Garnier, Sutton1999, Williams1992}:
\begin{equation}\label{eq:GradJ}
    \nabla_{\theta}J(\theta) = \mathbb{E}_{\tau \sim \pi_{\theta}} \left[\sum \limits_{{t=t_0}}^T \nabla_{\theta} \text{log}(\pi_{\theta}(a_t|s_t))R(\tau) \right]~.
\end{equation}
In this stochastic approach, the ascent of $J$ consists in first sampling the policy on different trajectories, then computing an average of the cumulative rewards obtained on these trajectories (the average thus replaces the expectation in Eq.~(\ref{eq:GradJ})), and finally adjusting $\theta$ with Eq.~(\ref{eq:updateThetaRec}), in order to progressively enhance the cumulative reward.

In the deterministic approach (deterministic policy-gradient algorithm, or DPG algorithm~\cite{Silver2014DeterministicPG}), the environment is still represented with a stochastic process (MDP), but the policy is considered to be deterministic. To keep in mind the difference, the deterministic policy is now represented by $\mu_{\theta}$. Replacing $\pi_{\theta}$ with $\mu_{\theta}$, the objective of the deterministic policy gradient algorithms is to apply the same principle as in the stochastic case. The mathematical formulation of the deterministic policy gradient is beyond the purpose of the present work. One may refer to the original work of \citet{Silver2014DeterministicPG}, and bear in mind that with a deterministic approach it is not needed to sample over a range of actions and states, but only over a range of states, as the actions are deterministically deduced.

\subsection{Limitations of the classical reinforcement-learning methods}

Classical RL methods, i.e. algorithms that use simple function approximators such as matrices, polynomials or other simple functions, may encounter limitations. Historically, applications of value-based methods to real-world problems~\cite{Tesauro1989APN, Crites1995ImprovingEP} have rapidly been overtaken by faster policy-based methods (see, e.g. the policy gradient method~\cite{Sutton2000}), which quickly proved their efficiency and applicability to various real-world tasks~\cite{Bagnell2001_appliRL,Ng2004_appliRL,PetersSchaal2006_appliRL}.

Through the idea of experience replay (see the original work of \citet{Lin1992}), value-based RL methods have known a revival (e.g., Ref.~\onlinecite{Riedmiller2007} for an application). The experience replay technique largely shrinks the number of interactions between the agent and the environment and thus accelerates these methods. Nevertheless, both the value-based and policy-based RL methods remain limited to problems with low state and action spaces~\cite{2015Mnih,Degris2012b}.

\section{Deep reinforcement learning}\label{sec:DRL}
Deep reinforcement learning (DRL) methods consist in the combination of neural networks and classical RL algorithms. Neural networks are known to be universal function approximators~\cite{Hornik1989MultilayerFN}, able to treat high-dimensional and non-linear problems, and thus to overcome the limitations set by the simple function approximators used in traditional RL. This means that, given a sufficient amount of example data representing a well behaved (typically, continuous) function, and a large enough ANN, the ANN can represent the function with arbitrary precision, i.e. ANNs large enough are dense in the space of continuous functions. Their application in reinforcement learning led to breakthroughs in various fields as they largely enhanced the performances of the previous methods in multi-input multi-output (MIMO) problems. To cite a few, the DRL algorithm developed by \citet{2015Mnih} obtained super-human levels in Atari games, and the one created by \citet{Silver2016} won against the European champion of Go five times in a row (to contextualize this achievement, no algorithm was able to beat a professional human player in this particularly complex board game before). Eventually, the PPO algorithm implemented by \citet{2019Rabault-al} successfully led to the first control of a flow by DRL.
\\

In this section, a short introduction to artificial neural networks is first provided. An overview of the main DRL methods -- presented in light of their RL precursors -- is then suggested. Within these methods, the most recent and encouraging developments are presented, as well as the challenges they face and try to overcome.

\subsection{Artificial neural networks}

\begin{figure}[!ht]
    \centering
    \includegraphics[width=0.5\textwidth]{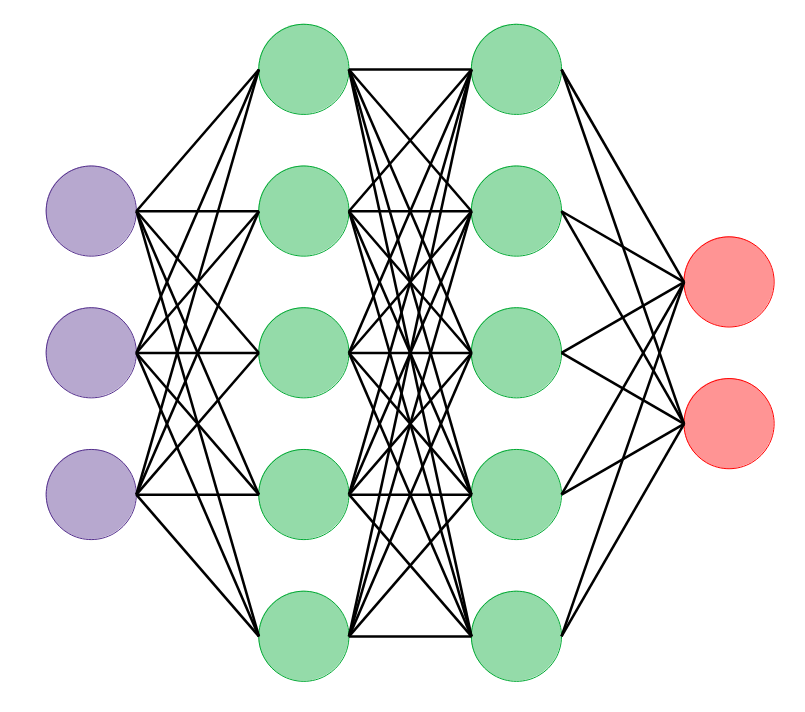}
    \caption{Schematic of a fully-connected artificial neural network. Purple denotes input, red output and green the hidden layers.}
    \label{fig:schemeFCNN}
\end{figure}

Artificial neural networks (ANNs, or NNs) are function approximators, based on connections between singular units, called neurons. A neuron is an entity characterized by three features : a set of weights $w = (w_1, ..., w_n)$, with $n\in \mathbb{N}^*$ corresponding to the number of inputs connected to the neuron, a bias $b$ ($b \in \mathbb{R}$), and the activation function $\sigma$, an hyper-parameter chosen at the conception of the neural network. Taking a vector $x = (x_1, ..., x_n)$ as an input, the neuron returns the scalar $\sigma(w.x + b)$. Neurons are usually organized in successive layers ; we then talk about deep NNs (DNNs). Each neuron of a layer is connected to one or more neurons of the precedent and next layers. When each neuron of each layer is connected to all the neurons of the precedent and of the next layer, the NN is said fully connected. FCANNs (fully connected artificial neural networks) are commonly used in DRL. A schematic of FCANN is given in figure \ref{fig:schemeFCNN}. This illustrative example is constituted of a layer of two output neurons (red disks), and two layers of five neurons (green disks). The first layer on the left of the figure is not constituted of neurons: the purple disks correspond to the inputs (here of size three). They are conventionally represented similarly to neurons but simply represent a scalar input (or, equivalently, a neuron with $n=1$, $b=0$, $w_1=1$ and $\sigma = \text{id}_{\mathbb{R}}$). The black lines correspond to the connections between the neurons.

The training of an ANN consists in tuning all its degrees of freedom, i.e tuning the biases and weights of all the neurons so that the ANN accurately represents a goal function known through examples. To do so, these parameters are commonly tuned step by step, by following, e.g., the back-propagating algorithm~\cite{LeCun2015}.

ANNs can be used in RL algorithms for their ability to estimate complex non-linear functions~\cite{Hornik1989MultilayerFN}. The number of NNs and the functions they approximate within the RL algorithm define several different DRL methods, which we analyze and compare in the following paragraphs.

\subsection{Value-based methods}

\subsubsection{Deep Q-learning} \label{sec:DQL}
Artificial neural networks are commonly known for their ability to deal with high-dimensionality and non-linearity, following their universal approximator properties, as previously discussed. Therefore, instead of filling a Q-table with the values of \textbf{$Q^{\pi}$} -- as it was done in value-based RL --, ANNs may be appropriate to directly approximate $Q^*$.
\\

The approximation of $Q^*$ by the neural network is represented by $Q_{\theta}$, with all the weights and biases of the neural network represented by the set $\theta$. The objective of the training phase of the neural network is to tune $\theta$ in order to achieve the best approximation of $Q^*$. To do so, a loss function is defined as the mean-squared error of the Bellman equation (Eq.~(\ref{eq:bellmanOptimallyCond})):

\begin{eqnarray}\label{eq:LossFunc_DQL_noApprox}
    L(\theta) = \mathbb{E}_{s,a}\bigg[\frac{1}{2}\Big \lbrace \mathbb{E}_{s'}\left[r(s,a) + \gamma \text{max}_{a'}Q_{\theta}(s',a')\right]
    - Q_{\theta}(s,a) \Big \rbrace^2 \bigg] \nonumber\\
    = \mathbb{E}_{s,a, s'}\left[\frac{1}{2}\left\lbrace r(s,a) + \gamma \text{max}_{a'}Q_{\theta}(s',a') - Q_{\theta}(s,a) \right\rbrace ^2 \right] \nonumber\\
    + \mathbb{E}_{s,a,s'}\left[\frac{1}{2}\mathbb{V}_{s'}\left[r(s,a) + \gamma \text{max}_{a'}Q_{\theta}(s',a')\right] \right].
\end{eqnarray}

The variance of $r(s,a) + \gamma \text{max}_{a'}Q_{\theta}(s',a')$ being independent of the weights $\theta$, the second term is usually ignored~\cite{2015Mnih}, and the loss function limited to:
\begin{equation}\label{eq:LossFunc_DQL_Approx}
    L(\theta) = \mathbb{E}_{s,a, s'}\bigg[\frac{1}{2}\Big\lbrace r(s,a) + \gamma \text{max}_{a'}Q_{\theta}(s',a') 
    - Q_{\theta}(s,a) \Big\rbrace^2 \bigg]~.
\end{equation}

More precisely, at each time-step i:
\begin{equation}\label{eq:LossFunc_DQL_iter}
    L_i(\theta_i) = \mathbb{E}_{s,a,s'}\bigg[\frac{1}{2}\Big\lbrace r(s,a) + \gamma \text{max}_{a'}Q_{\theta_{i-1}}(s',a')
    - Q_{\theta_i}(s,a) \Big\rbrace^2 \bigg]~.
\end{equation}

The objective of the neural network is to minimize the loss function, thus to modify $Q_{\theta_i}$ to get it closer to the target $r(s,a) + \gamma \text{max}_{a'}Q_{\theta_{i-1}}(s',a')$, computed with the weights $\theta_{i-1}$ of the precedent update. To this end, the update:
\begin{equation}
    \theta_{i+1} \longleftarrow \theta_i - \alpha \nabla_{\theta_i}L_i(\theta_i)~,
\end{equation}
is applied (with $\alpha$ a positive parameter) and should make $Q_{\theta}$ converge to $Q^*$.
\\

However, a naive coupling -- such as the one mentioned above -- between an ANN and a RL algorithm would be compromised by two issues that may lead to divergence. First, the process of learning from sequences of actions and observations on-policy induces correlations. Indeed, the algorithm is potentially biased if it solely learns from the successions of tuples $(s_t, a_t, s_{t+1},r_t)$ that occur during the running episode (on-policy), because these tuples are generally not independent within an episode, which has a tendency to make the NN training fail. Furthermore, a minor modification of the value of $Q$ may change significantly the data distribution and thus may lead to an unstable algorithm.

In order to tune these two biases, \citet{2013Mnih} suggest the integration of a replay buffer in their algorithm. Filled  with a certain amount of tuples $(s_t, a_t, s_{t+1},r_t)$ that occurred in past episodes, the replay buffer enables the ANN to have more uncorrelated input data and, simultaneously, a more stabilized data distribution, which is required for a correct learning process. By adapting the experience replay technique suggested in Ref.~\onlinecite{Lin1992}, \citet{2013Mnih} thereby propose the Deep Q-Learning (DQL) algorithm, which masters complex control policies in Atari games. \citet{2015Schaul} extend the idea of replay buffer by suggesting to optimize its filling, by favouring less frequent actions instead of a uniform process, as the learning phase could benefit from these unusual events.
\\

Even with the DQL algorithm, a major restriction remains, due to the fact that the neural network is chasing a moving target. As developed above, the ANN tries to reach the Bellman optimally condition (Eq.~(\ref{eq:bellmanOptimallyCond})) by minimizing the loss function $L$. To do so, it modifies $Q_{\theta_i}$ to get it closer to the target $r(s,a) + \gamma \text{max}_{a'}Q_{\theta_{i-1}}(s',a')$, but the target itself changes at each timestep $i$. This issue induces instability and divergence in many situations, notably when the estimation of Q is provided by a nonlinear function approximator~\cite{Tsitsiklis97}, such as a FCANN. This instability thus highly reduces the algorithm robustness.

\subsubsection{Deep Q-networks} \label{sec:DQN}
Developed by \citet{2015Mnih}, Deep Q-networks (DQNs) are an evolution of DQL. The DQN method mainly improves the ability of the neural network to treat high-dimensional environment spaces and suppresses the issue of divergence detailed above, by providing an innovation to the DQL algorithm.

The problem of divergence is solved by exploiting a second ANN, a 'slow-reacting clone' of the original one. Let us define the loss function with Eq.~(\ref{eq:LossFunc_DQL_Approx}). In DQL, the algorithm tries to equalize the target $r(s,a) + \gamma \text{max}_{a'}Q_{\theta}(s',a')$ with $Q_{\theta}(s,a)$, by using a single neural network. In DQN, an ANN is applied for the computation of $Q_{\theta}$ and the 'slow-reacting' copy is applied for the estimation of the target. Denoting $\theta'$ as the set of weights and biases of this second network, the loss function is then defined by:
\begin{equation}\label{eq:LossFunc_DQN}
    L(\theta) = \mathbb{E}_{s,a, s'}\bigg[\frac{1}{2}\Big\lbrace r(s,a) + \gamma \text{max}_{a'}Q_{\theta'}(s',a')
    - Q_{\theta}(s,a) \Big\rbrace^2 \bigg]~.
\end{equation}

The strategy suggested by \citet{2015Mnih} is to regularly equalize $\theta$ and $\theta'$. If such an equalization is made at each update of $\theta$, then the method becomes identical to DQL. Different strategies have been developed to correlate $\theta$ and $\theta'$. In Ref.~\onlinecite{2015Lillicrap}, the authors develop a different method (called DDPG and detailed below), but use the same idea, and propose the iterative process $\theta' \longleftarrow \tau \theta + \left(1-\tau\right)\theta$, with $\tau \ll 1$. In the original DQN algorithm~\cite{2015Mnih}, $\theta'$ is actualized one time over N updates of $\theta$ ($N \in \mathbb{N}^* \gg 1$). These strategies with two ANNs thus eliminate the instability by slowing down the movements of the target.
\\

Dueling DQN~\cite{Wang2015DuelingNA} is an evolution of DQN in which the neural network is split into two streams, one aimed at evaluating the advantage function, while the other one computes the state value function $V^{\pi}$. The Q-function is then estimated by combining the approximations of $A^{\pi}$ and $V^{\pi}$ (in a non-trivial manner). The advantage function depends on the actions chosen whereas the state value function does not. By taking advantage of these two functions and their different dependencies on the actions, the dueling DQN algorithm overcomes the DQN method particularly in situations where there are many similar-valued actions.

\subsubsection{Double deep Q-learning and double deep Q-networks}\label{sec:doubleDQL_DQN}
Double DQL and double DQN~\cite{VanHasselt2010, 2015VanHasselt} have been developed in order to solve the issue of overestimation happening in DQL and DQN, a problem partially unknown or ignored up to then. The overestimation in these algorithms is due to the fact that in the computation of the target, the same neural network is used both for the selection of the action and for its evaluation. Indeed, in DQL, one can define the target $Y$ as:
\begin{eqnarray}\label{eq:targetDQL}
    Y(s,a) = r(s,a) + \gamma \text{max}_{a'}Q_{\theta}(s',a') \nonumber\\
    = r(s,a) + \gamma Q_{\theta}(s', \text{argmax}_{a'}Q_{\theta}(s',a'))~,
\end{eqnarray}
and likewise in DQN:
\begin{eqnarray}\label{eq:targetDQN}
    Y(s,a) = r(s,a) + \gamma \text{max}_{a'}Q_{\theta'}(s',a') \nonumber\\
    = r(s,a) + \gamma Q_{\theta'}(s', \text{argmax}_{a'}Q_{\theta'}(s',a'))~.
\end{eqnarray}

In both cases, the neural network that estimates $Q^*$ is used to choose $a'$ and then to apply $Q^*(s',a')$, which can lead to an overestimation of the expected cumulative reward. Not only is overestimation quite common, but it can also adversely affect the performances of the algorithm~\cite{2015VanHasselt}. For the DQL algorithm, a new method with a second network is presented in Ref.~\onlinecite{2015VanHasselt}, inspired by previous works~\cite{VanHasselt2010}. In double DQL, two networks are exploited so that one focuses on the choice of the action, and the other one on the evaluation. Hence, the target becomes:
\begin{equation}
    Y(s,a) = r(s,a) + \gamma Q_{\theta^{bis}}(s', \text{argmax}_{a'}Q_{\theta}(s',a'))~,
\end{equation}
with $\theta^{bis}$ the set of weights and biases of the new neural network, used for the evaluation. The two networks regularly invert their roles, so that $\theta$ and $\theta^{bis}$ are similarly updated. The double DQN algorithm does not require any other network, a second one being already exploited within the DQN method. Indeed, \citet{2015VanHasselt} suggest:
\begin{equation}
    Y(s,a) = r(s,a) + \gamma Q_{\theta'}(s', \text{argmax}_{a'}Q_{\theta}(s',a'))~.
\end{equation}

According to \citet{2015VanHasselt}, the performances of the value-based DRL methods are enhanced, even the already robust performances of the DQN algorithm in Atari games~\cite{2015Mnih}.

\subsubsection{Challenges and openings}

The DQN algorithm has first been developed to stabilize the DQL method, and notably enhanced its efficiency to deal with large neural networks. Thanks to that, Deep Q-Networks have shown great results with high-dimensional observation spaces~\cite{2015Mnih}. But DQN or even double DQN algorithms are not easily able to deal with high-dimensional action spaces. Indeed, as they seek $\text{argmax}_a Q_{\theta}(.,a)$ at each timestep, increasing the dimensionality of the actions space highly reduces their performances. Some methods have been developed to overcome this issue, making continuous action spaces possible. Based on a combination of policy gradient and Q-function, they also offer larger possibilities, as they may consider either off-policy or on-policy strategies, whereas the methods mentioned above are limited to off-policy strategies~\cite{Mnih2016}, due to the replay buffer. These promising policy-based methods are presented in the next section.

\subsection{Policy-gradient and actor-critic methods}

\subsubsection{Deep policy gradient}

Amidst the current most promising DRL methods, the second main contender is the family of actor-critic methods. These are obtained by combining elements of Q-learning and policy-gradient methods, which we will discuss now.

The deep policy gradient algorithm is a policy-based method, i.e it directly optimizes the policy (see section \ref{sec:PB_methods}). Explicitly, an ANN approximates the policy $\pi_{\theta}$, with $\theta$ representing its set of weights and biases. The loss function is defined by:
\begin{equation}\label{eq:LossFunc_deepPG}
    L(\theta) = -\mathbb{E}_{\tau \sim \pi_{\theta}} \left[\sum \limits_{{t=t_0}}^T \text{log}(\pi_{\theta}(a_t|s_t))R(\tau) \right]~,
\end{equation}
so that $\nabla_{\theta}L(\theta) = - \nabla_{\theta}J(\theta)$ (see Eq.~(\ref{eq:GradJ}) for $\nabla_{\theta}J$). This expected value is approximated by an average over a set of trajectories. $R(\tau)$ is commonly replaced with the advantage function $A^{\pi_{\theta}}$, as it reduces the variance of the expectation~\cite{Schulman2015}
and thus reduces the number of trajectories necessary to well average this expectation. Hence, in the deep policy gradient algorithm, the loss function:
\begin{equation}\label{eq:LossFunc_deepPG_advantage}
    L(\theta) = - \mathbb{E}_{\tau \sim \pi_{\theta}} \left[\sum \limits_{{t=t_0}}^T \text{log}(\pi_{\theta}(a_t|s_t))A^{\pi_{\theta}}(s_t, a_t) \right]~,
\end{equation}
is usually considered. Nonetheless, as the expectation is averaged over a set of trajectories at the end of an episode, the algorithm does not differentiate between the most accurate actions and the worse, but only computes the reward obtained from the whole trajectory. Actor-critic methods adapt the deep policy gradient algorithm to deal with this issue.

\subsubsection{Advantage actor-critic methods}

Deep policy gradient methods are limited as they do not differentiate positive from negative actions. To clear it up, actor-critic methods rely on a combination of policy-based and value-based methods, as one neural network (the actor) approximates $\pi_{\theta}$, the policy by which the actions are chosen, and a second one (the critic) computes the Q-function $Q_{\theta'}$, in order to evaluate the goodness of the action taken by the agent. With the same prospect of variance reduction, $Q_{\theta'}$ is often replaced with $A_{\theta'}$, an estimation of the advantage function. The method is then called Advantage Actor-Critic (A2C)~\cite{SuttBarto2018}.
\\

One may refer to the review of \citet{2021Garnier} for explicit implementations of the vanilla deep policy gradient and A2C algorithms. Many variations have been developed, such as the asynchronous A2C (A3C) suggested in Ref.~\onlinecite{Mnih2016}. The latter was presented to have similar - if not better - performances compared to the DQN algorithm in Atari games, and additionally to be able to treat continuous action spaces, which was a reason for the development of policy-based methods. Indeed, as an $argmax_{a\in A}()$ function is not computed anymore, these methods enable one to tremendously increase the actions space $A$, up to a continuous space in some algorithms.

\subsubsection{The deep deterministic policy gradient method}

As explained above, a major limit of DQN and other value-based methods is their lack of robustness with high-dimensional action spaces. The deep deterministic policy gradient (DDPG) algorithm was first developed by \citet{2015Lillicrap} to solve this specific issue. It is a model-free off-policy actor-critic method, inspired by the Deterministic Policy Gradient (DPG) algorithm presented in Ref.~\onlinecite{Silver2014DeterministicPG} (see section \ref{sec:PB_methods}). Contrarily to the previously presented DRL methods, the DDPG (like the DPG) algorithm considers a deterministic policy mapping, and not stochastic. Thus the actions space is explored thanks to a perturbative method:
\begin{equation}
    a_t = \mu_{\theta + N_1} \left( s_t \right) + N_2~,
\end{equation}
with $N_1$ and $N_2$ two different noisy processes, $\theta$ the set of weights and biases of the actor, and $a_t$ the action taken at timestep $t$ in state $s_t$ according to the deterministic policy $\mu$. The critic computes the optimal Q-function similarly to the DQN algorithm, but replaces the $argmax_{a\in A}()$-seeking with a gradient over $\mu$ (see Ref.~\onlinecite{2015Lillicrap}), and thus enables one to leverage high-dimensional and continuous action spaces.

\citet{2019Bucci} succeeded in stabilizing the dynamics of a chaotic system governed by the one-dimensional Kuramoto-Sivashinsky equation thanks to a DDPG controller, while more recently \citet{2022Hasegawa} applied this algorithm to a flow control problem (the reduction of the skin friction drag in a channel flow). Both of these examples illustrate the promising performances of the DDPG method with high-dimensional action spaces.

\subsubsection{Proximal policy optimization}

The proximal policy optimization (PPO) algorithm is an on-policy actor-critic method, developed by \citet{2017Schulman} in order to deal with the lack of robustness of the DQN~\cite{2015Mnih} and 'vanilla' policy gradient~\cite{Mnih2016} methods, and simplify the efficient-but-complex trust region policy optimization (TRPO) algorithm~\cite{Schulman2015}, not discussed in the present review.

Apart from the particular DDPG algorithm, the policy gradient and actor-critic methods usually aim at optimizing the loss function defined in Eq.~(\ref{eq:LossFunc_deepPG_advantage}), now denoted by $L^{PG}(\theta)$. To do so, these algorithms proceed over several epochs of gradient ascent, i.e they alternate a first phase where they fill a batch of samples (a sample being a sequence $(s_t, a_t, s_{t+1}, r_t)$ for instance), and a second phase in which the expectation of the gradient of $L^{PG}(\theta)$ is averaged over the batch, and then the neural network parameters $\theta$ updated following the direction of this estimation. The idea behind the TRPO and PPO algorithms is identical, but $L^{PG}$ is replaced with other functions. For instance, the surrogate loss:
\begin{eqnarray}\label{eq:L_clip}
    L^{\textsc{Clip}}(\theta) = \mathbb{E}_{\tau \sim \pi_{\theta}} \Bigg[\sum \limits_{{t=t_0}}^T \text{min}\big\lbrace r_t(\theta)A^{\pi_{\theta}}(s_t, a_t); \nonumber\\
 \text{clip}(r_t(\tau), 1 - \epsilon, 1+\epsilon)A^{\pi_{\theta}}(s_t, a_t) \big\rbrace \Bigg]~,
\end{eqnarray}
is suggested in Ref.~\onlinecite{2017Schulman} for the PPO algorithm, with:
\begin{equation}
    r_t(\theta) = \frac{\pi_{\theta}(a_t|s_t)}{\pi_{\theta^{\text{old}}}(a_t|s_t)}~,
\end{equation}
being a factor that penalizes an update of $\theta$ that would induce an important modification of the policy (comparatively to the situation before the update, with the previous set of NN parameters $\theta^{\text{old}}$). $\epsilon$ is a parameter chosen between 0 and 1, and $\text{clip}(r_t(\tau),~1~-\epsilon,1+\epsilon)$ returns $r_t(\tau)$ if its value is in $[1 - \epsilon; 1+\epsilon]$,  $1+\epsilon$ if it is greater than this limit, and $1-\epsilon$ otherwise. In the TRPO algorithm, the objective function is similarly penalized according to the size of the update~\cite{Schulman2015}. The PPO algorithm is a simplification of this previous work as it only considers a first order optimization process.  Another variant of loss function is proposed in Ref.~\onlinecite{2017Schulman}, using a KL-penalty~\cite{KakadeLangford2002}.
\\

After having tuned $\epsilon = 0.2$, \citet{2017Schulman} observe that the PPO algorithm outperforms most of the previous online policy gradient algorithms (A2C and vanilla PG algorithms~\cite{Mnih2016}, a combination of A2C and trust region algorithm~\cite{2016Wang}, the TRPO method~\cite{Schulman2015} and the cross-entropy method (CEM)~\cite{SzitaLorincz2006}), on almost all the continuous control environments.

Hence, the PPO algorithm is one of the most promising modern DRL methods, and has already successfully been applied to a variety of problems (see, e.g., Ref.~\onlinecite{2018Peng}). It is often considered as the 'state-of-the-art' for continuous control~\cite{2021Ren_hydroStealth}. To cite an application in flow control, \citet{2019Rabault-al} used a PPO algorithm for performing the first active flow control combining CFD simulations and deep reinforcement learning, in the standard example of the cylinder immersed into a 2D flow, and \citet{2021Ren_AFCturb} extended the previous work to a weakly turbulent regime.
\\

Variations on the basis of the original PPO algorithm now flourish, such as PPO-1~\cite{2021Ghraieb}. PPO-1 is a degenerate~\cite{2020Viquerat} version of the PPO method, developed for open-loop control. The algorithm runs multiple simulations in parallel, shuffles the data coming from all the environments and fills several mini-batches with these mixed data, then uses these mini-batches sequentially to update the NN, and eventually repeats the sequence by launching multiple parallel simulations again, etc. Another variant is the PPO with Covariance Matrix Adaptation (PPO-CMA)~\cite{2020Hamalainen}, an algorithm able to dynamically expand the variance of the exploration policy to speed up the seeking phase, and then to shrink it when the algorithm gets close to the global optimum. PPO is stable but sometimes prematurely reduces the exploration variance, and then may converge to local optima, hence the development of this algorithm. The reader may also refer to the Maximum a posteriori Policy Optimization (MPO) algorithm~\cite{2018Abdolmaleki1} and extensions~\cite{2018Abdolmaleki2} for similar motivations.

A recent variation of PPO-CMA, called AS-PPO-CMA~\cite{Paris2022_bis}, aims at reducing the number of actuators while limiting the loss induced in the cumulative reward. This development goes along with the prospect of increasing the complexity of the problems handled by DRL, where an effective reduction of the actions space may be critical.

\begin{figure}
    \centering
    \includegraphics[width=0.5\textwidth]{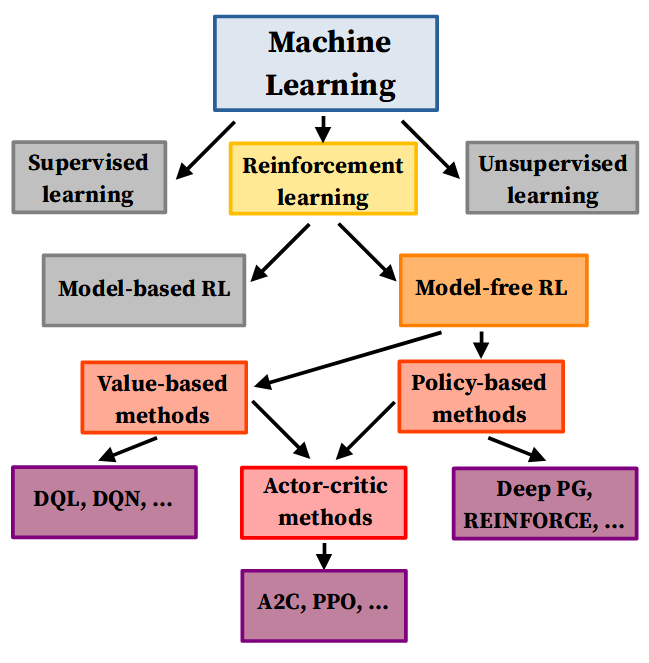}
    \caption{Non-exhaustive classification of machine learning methods, with focus on the reinforcement-learning-based techniques.}
    \label{fig:ML_tree}
\end{figure}

\begin{figure*}
    \centering
    \includegraphics[width=0.8\textwidth]{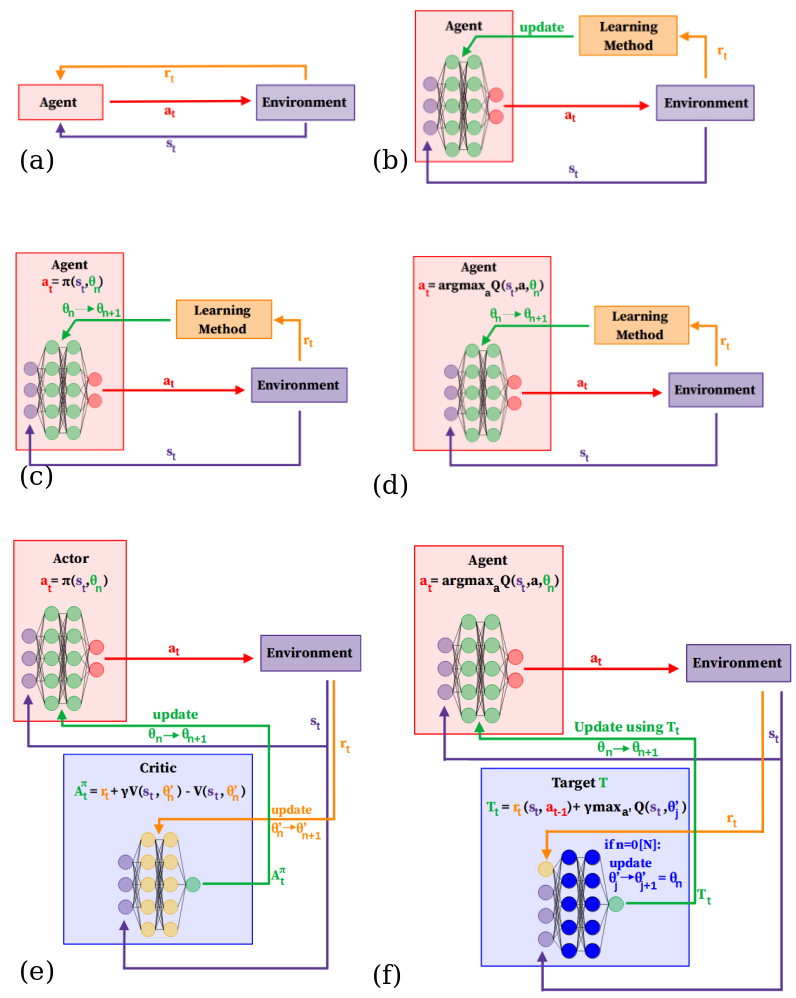}
    \caption{Comparison between (a) RL, (b) DRL, (c) policy-based DRL, (d) value-based DRL, (e) advantage actor-critic DRL and (f) DQN (simplified) architectures.}
    \label{fig:All_architectures}
\end{figure*}

For comparison, a non-comprehensive overview of RL methods is suggested in figure \ref{fig:ML_tree}, while schematics representing the main different DRL architectures are gathered in figure \ref{fig:All_architectures}.

\subsection{Parallelization}\label{sec:Parallelization_in_DRL}

Multiple processes have been developed to reduce the computational cost of DRL algorithms, in order to then consider more complex tasks. The ones gathered by the idea of parallelization are presented in this section.
\\

First, when it comes to fluid dynamics problems that combine CFD simulations with DRL algorithms, one may consider the parallelization of the simulations themselves. This subject has a long history~\cite{Simon1992, GROPP1990}, and well-implemented parallel CFD solvers nowadays reach limitations correlated to the number of nodes~\cite{GROPP2001}, hence limiting the gain of time with parallel CFD simulations.

Consequently, as the computational cost due to the simulations may remain important even with parallel CFD solvers, the parallelization of the DRL algorithms is of great interest. Speedups would enable one to study more challenging flow configurations, which are the main objective of the current researches in the field. \citet{2019RabaultKuhnle} consider the simple bi-dimensional benchmark presented in Ref.~\onlinecite{2019Rabault-al} and adapt the PPO algorithm to parallelization. Their idea is to run independent episodes on several parallel environments, so that the collection of data used for the training phase is parallelized. Indeed, they prove that if one runs $N$ episodes sequentially between two updates of the neural network, running these $N$ episodes in parallel is equivalent, and thus divides the computational cost by approximately $N$. Results are conclusive, the computational time being divided by twenty in their most optimal implementation. This is of great interest, as problems with higher complexity usually need to increase the size of the batch and the number of episodes between updates of the NN. Increasing the complexity would therefore no longer induce an increase of the computational cost, if additional parallel environments were added accordingly. The idea of PPO-1~\cite{2021Ghraieb} is to achieve multiple simulations in parallel, then to fill several mini-batches with the shuffled data, and eventually to sequentially use the mini-batches to update the NN. The parallelization process is thus very similar to the one of \citet{2019RabaultKuhnle}.

Massively distributed architectures on multiple GPUs have been implemented to accelerate DRL methods, such as DQN with the Gorila framework~\cite{2015Nair}. Multiple agents act in parallel on identical copies of the environment, have their own replay buffer and compute the loss function similarly to the DQN process (Eq.~(\ref{eq:LossFunc_DQN})). A central actor merges the information from all the agents, and then enforces the same ANN update to each of them. With 100 parallel agents, Gorila outperformed DQN on many Atari games, with computations approximately twenty times faster. A similar process has also been suggested by \citet{Chavez2015}, whereas \citet{Mnih2016} implemented the same principle but with multiple CPU threads on a single machine, thereby outperforming the previous methods while offering the possibility to have stable value-based and policy-based, on-policy and off-policy efficient strategies. Indeed, an important improvement brought by the parallel agents is that it stabilizes value-based methods even without any replay buffer~\cite{Mnih2016}, thus allowing more efficient on-policy methods.
\\

Eventually, \citet{belus2019} exploit for the first time the translational invariance and the locality of the control problem in a DRL algorithm, in order to enhance its efficiency. Their work is not a parallelization process but also enables one to reach more complex problems, and it will be more extensively discussed in the last section.
\\

\section{Deep reinforcement learning for flow control} \label{sec:DRL_and_flow_control}
\subsection{Introduction passive and active control}

Flow control strategies are usually separated in two categories~\cite{GadElHak96}: passive and active methods. The former include, e.g., riblets and vortex generators~\cite{COLLIS2004} and other LEBUs (large-eddy-break-up devices)~\cite{Alfredsson2018_LEBUsReview}. They do not need a supply in energy nor a feedback from the environment. Their robustness, low manufacturing costs and global simplicity have notably enabled them to be applied in real-world systems, contrarily to the active methods which are often complex to implement concretely. But the active methods present the advantage to be generally more efficient, and in a wider range of operating conditions. For instance, while riblets used for drag reduction in turbulent boundary layers can perform a 5 to 9\% drag reduction~\cite{Bechert1989}, modern active methods can perform reductions twice as much important, or even more (see Ref.~\onlinecite{COLLIS2004}, or Ref.~\onlinecite{2022Vinuesa} for a recent review on flow control in turbulent boundary layers).

Passive methods are intrinsically open-loop as they do not consider the evolution of the surrounding environment. However, active methods can be either open-loop or closed-loop. Open-loop active methods, by not updating the actuators with a potential feedback of the environment state, are sub-optimal. But having a good understanding of the open-loop behavior is almost a prerequisite to the further development of closed-loop actuators~\cite{COLLIS2004}. For instance, actuators used to control flow separation have first proved to be efficient in periodic regimes~\cite{GREENBLATT2000, Seifert1996} and paved the way to modern closed-loop actuators (see Ref.~\onlinecite{2019Rabault-al} for an example of closed-loop flow control using synthetic jets).
\\

For historical and comprehensive reviews on active and passive methods, the reader may refer to the works of \citet{GadelHak1998FlowC}, \citet{MoinBewley1994} and \citet{LumleyBlossey1998}, and to the review of \citet{2022Pino} for the use of machine learning methods in flow control. The objective of this section is to give a general overview of the last advances in active flow control (AFC), and more particularly in the application of the DRL methods to that field. The current challenges and upcoming milestones in the combination of DRL and AFC are highlighted.

\subsection{Active flow control}

Active flow control has grown at the frontier of various traditional fields, including fluid mechanics, control theories and machine learning~\cite{BEWLEY2001}. Thus, it has inherently been confronted to terminology conflicts~\cite{Sutton1992_terminologyConflict, Bersini96, BEWLEY2001, Recht2019, Nian2020}, but has developed fruitful hybrid methods during the past two decades. This interdisciplinary effort has indeed been carried out with the prospect of many promising applications~\cite{2022Pino, BruntonNoack2015}, such as delaying the transition past airfoils/aircraft wings~\cite{Kuhn2011, batikh:hal-01820331}, reducing the drag coefficient past bluff bodies (see, e.g. Ref.~\onlinecite{2021Ren_hydroStealth}) or controlling convective heat transport~\cite{2020Beintama}, to cite a few.

\begin{figure}
    \centering
    \includegraphics[width=0.5\textwidth]{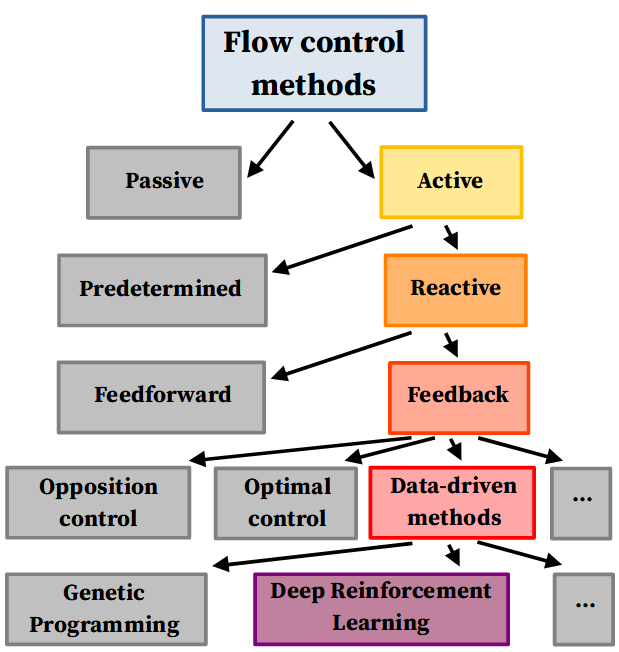}
    \caption{Non-exhaustive classification of flow-control strategies.}
    \label{fig:AFC_tree}
\end{figure}

Contrarily to the passive methods, the active control of a flow needs a supply in energy. In addition, the data used for the control can be provided by experimental results or CFD simulations. These data can be used during the control sequence (reactive control) or only before (predetermined control). As feed-forward methods are not considered in the present review, active reactive feedback methods will simply be called feedback control thereafter (see Fig.\ref{fig:AFC_tree}). In terms of terminology reconciliation, open-loop and closed-loop (D)RL methods applied to the control of a flow are respectively assimilated to predetermined and feedback control. The closed-loop DRL methods are thus an active reactive feedback data-driven strategy (see Fig.\ref{fig:AFC_tree}).

\subsubsection{Predetermined control}
The predetermined control of a flow is regardless of its instantaneous state. The control sequence is determined in advance, and then applied without any feedback. It facilitates the control, since no sensors are needed and no feedback loop has to be implemented~\cite{2022Hasegawa}.

In a numerical approach, \citet{Jung92} and then \citet{QuadrioRicco2004} have enforced spanwise oscillations at the walls of a fully developed turbulent channel, and computed reductions of the turbulent drag of the order of 40\%. Several studies~\cite{Atzori1,Atzori2,Fahland} have also considered predetermined control based on blowing and suction in turbulent wings, also leading to over $10\%$ improvement in aerodynamic efficiency. Alongside these direct numerical simulation approaches, experimental versions of the spanwise oscillating walls have been implemented~\cite{Laadhari94, Choi2002, KarniadakisChoi2003}, and important but lower values (25-35\%) of drag reduction were observed. \citet{QuadrioRicco2004} also reported a net energy saving of 7\% (taking into account the power necessary for the wall oscillations), hence highlighting that even without a feedback algorithm, sensors or complex actuators, a predetermined control method can be efficient. 

Other predetermined methods have been tested in the field of flow control. One may refer to Refs.~\onlinecite{Min_al2006,2010Lieu} for similar but streamwise blowing and suction, to Refs.~\onlinecite{Kametani2011,Kuhn2011} for uniform blowing and suction, and to Ref.~\onlinecite{Kuhn2011_part2} for a periodic blowing on a wing flap (high-lift device).  All these studies illustrate the efficiency of actuators used in a steady or cyclic manner with open-loop algorithms. Nevertheless, as they do not take into account the evolution of the dynamics in the flow, their range of efficiency remains limited. \citet{QuadrioRicco2004}  explain that the input parameters of the oscillations depend on the Reynolds number for instance. In complex turbulent flows, predetermined methods thus encounter limitations similar to the passive methods.

\subsubsection{Feedback control}
Feedback control includes a closed-loop process to take into account the instantaneous evolution of the flow. It is sometimes called \textit{reactive} -- or even briefly \textit{active} -- control. The number of actuators, their degrees of freedom and the quantity of input data -- collected by one or more sensors -- can make these methods challenging. Within these methods, multiple processes can be distinguished. Opposition control methods~\cite{Choi1994}, optimal control theory (see, e.g., Ref.~\onlinecite{Sargent2000}), genetic programming~\cite{1992Koza}, reinforcement learning~\cite{sutton1988learning} and deep reinforcement learning~\cite{SuttBarto2018} applied to AFC problems are successively discussed in the present section, and compared in view of the milestones to reach. If well implemented, the feedback should enhance the efficiency of the control sequence, by adapting itself to the 'live' state of the flow. For instance, \citet{Ren2019} obtained an encouraging 20\% improvement of the control performances by applying a closed-loop process rather than any open-loop method in the AFC problem they were considering.

\begin{description}
    \item[Opposition control] has first been developed by \citet{Choi1994} to reduce the drag generated by the skin friction in a fully developed turbulent channel flow, by trying to annihilate the coherent structures in the boundary layer with a feedback process. These coherent structures play a preponderant role in the transport of turbulences~\cite{Cantwell1981OrganizedMI, Robinson1991CoherentMI}, and such a drag reduction would have numerous applications, since this type of wall-bounded flows are classical in engineering problems. The idea behind opposition control is to apply local blowing and suction at the wall according to a sensing detection plane, parallel to the wall and typically located at $h^+ = 15$ wall units from the wall~\cite{Hammond1998}. For instance, if a normal speed of $v(x,y, h^+)$ is detected at the position $(x,y,h^+)$ in the detection plane, a normal blowing (or suction, according to the sign of $v$) of $-\alpha v(x,y,h^+)$ is applied at the point $(x,y,0)$ of the wall (with $\alpha>0$ the linear coefficient). By controlling either the normal (as presented above) and/or spanwise velocities in such a simple feedback control, a skin friction reduction of approximately 20-25\% is obtained~\cite{Choi1994, Chung2011}. These methods are thus simple and show relatively efficient results, but they become unusable in more complex flows. Indeed, in the standard situation presented above, the relationship between the detected speed and the actuation is considered to be linear, and the optimal coefficient ($\alpha$) is found by trial and error. But this trial and error process is quite inefficient, even in the linear approximation~\cite{2022Hasegawa}. Additionally, the drag reduction rate obtained by opposition control in turbulent boundary layers decreases with increasing Reynolds numbers~\cite{Stroh2015}. Since then, the opposition control method is not aligned with the current objectives of controlling more sophisticated and turbulent flow conditions.
    
    \item[Optimal control theory] can efficiently optimize an input parameter, even with large degrees of freedom, by directly tackling the mathematical equations of the flow, and showed successful results in AFC in a weakly turbulent channel configuration~\cite{Bewley2001_optimalControlFluids}. In optimal control, the dynamical evolution of the system is explicitly developed until a certain time horizon, and then 'rewound' back to the initial state in order to solve the adjoint equations and compute a certain cost function. The optimal value of the parameter at stake is then tuned according to the cost function. As the non-negligible time horizon and the resolution of the adjoint equations require important computation time and resources, a sub-optimal theory as been developed~\cite{lee_kim_choi_1998, 2011HasegawaKasagi}, carrying the strategy of setting an arbitrarily small time horizon (thus suppressing the necessity to solve the adjoint equations~\cite{2022Hasegawa}) while limiting the deterioration of the optimization. But sub-optimal control solutions are limited. Indeed the more complex and turbulent the flow becomes, the more longer the time horizon needs to be, in order to have a good understanding of the dynamics of the system (for a further optimization of the parameter at stake). Additionally, the time horizon cannot be extended endlessly in optimal control, the adjoint equations being unstable beyond a certain threshold~\cite{WANG2014}. Consequently, in prospect of (actively) controlling flows with high Reynolds numbers, the optimal control theory may present non-negligible shortcomings.

    \item[Data-driven methods] are promising, since they could overcome the previous flaws encountered within the opposition and optimal control methods. They take various shapes but are gathered by the same preeminence of/need for consequent data.  

    \textbullet \quad The Bayesian optimization (BO) and Lipschitz global optimization (LIPO) processes are 'black-box' global optimization techniques~\cite{2022Pino}. Compared to the upcoming genetic programming and RL methods, they present the advantage to be less complex, as they are (respectively) linear and quadratic function approximators, and to require less computational time. In the classical BO method, the function to optimize (e.g. the cumulative reward) is modeled by a Gaussian process. Given a certain data set, one can thus obtain a probability distribution of the values taken by the function at stake. In the BO iterative process, a function suggesting where to sample next is also defined. The LIPO method relies on the same basis, but with a different surrogate function instead of the Gaussian process. One can refer to the work of \citet{2022Pino} for a more extensive and complete formulation of the BO and LIPO methods applied to AFC, and to the work of \citet{2021Blanchard_BOforAFC} for the first application of open-loop BO in AFC. 
    
    \textbullet \quad A genetic programming control algorithm is a nature-inspired feedback process that relies on the succession of populations~\cite{1992Koza}. A population of individuals (i.e. a set of control laws) is first generated, applied to the environment (i.e. a numerical simulation or experiment per control law is run), and then multiple processes (replication, cross-over, mutation amidst the 'individuals', i.e. the control laws) modify the previous population to produce a new one, and gradually optimize a function (e.g. the cumulative reward). Examples of successful GP applications in AFC can be found in Ref.~\onlinecite{gautier2015} for a separation control issue, in Ref.~\onlinecite{2017Li_GP_dragRed} for a drag-reduction-behind-a-car problem, in Ref.~\onlinecite{FanNoack2019} for the mixing optimization of a turbulent jet, in Ref.~\onlinecite{Hao2020} for an example of open-loop GP control, or more quantitatively in the review proposed by \citet{Noack2019_Book_EnhanceKnowledge}. The aforementioned applications used experimental setups to feed the closed-loop process with feedback data. \citet{Ren2019} present an innovative combination of CFD and GP-based AFC to suppress the vorticity induced by a vibrating cylinder. Their results outperform the overall open-loop methods by approximately 20\%, and thus arouse encouraging prospects.

    \textbullet \quad Eventually, the application of DRL in AFC has already shown promising results, and overtook the majority of the results obtained with previous methods. With the objective of increasing the complexity of the handled flow configurations, DRL methods appear to be the most relevant contender. As a consequence, a specific section is dedicated to these methods (see section \ref{sec:subsec_DRLinAFC}).
    
\end{description}

By applying PPO to laminar flow conditions ($Re = 100$), \citet{Tokarev2020} obtain performances comparable to the ones achieved with an adjoint-based method in Ref.~\onlinecite{FlinoisColonius2015}. However, in the situation of a wall-bounded channel flow, and even at a low friction Reynolds number (of about 180), the DRL algorithm developed by \citet{Guastoni2023} outperforms the opposition control method by over 20 percentage points, with drag reductions of, respectively, 46\% and 20\%. Furthermore, opposition and optimal control reach limitations when it comes to complex tasks such as controlling high-Reynolds turbulent flows, because the design of a relevant explicit control law becomes challenging, if not impossible. \citet{2022Pino} additionally illustrate that GP and DDPG outperform BO and LIPO on three showcases with progressive difficulty. Nonetheless, as the generality of the function approximator increases, the number of episodes required for the training phase grows in proportion. Therefore, it may worth combining GP and DRL algorithms with 'simpler' methods~\cite{2022Pino, 2019Li_al, Mendez2023}, or associating model-free DRL algorithms with environments partially modeled thanks to the physical understanding of the dynamics of the system (see, e.g., Refs.~\onlinecite{2021LiZhang,2021Qin} and the references therein). Compared to DRL, GP presents the advantage of performing better with fewer sensors~\cite{Castellanos2021}. However, DRL methods are inherently more suitable for multi-input multi-output problems than GP~\cite{2021Ren_hydroStealth}. The last section of this review will therefore be devoted to the application of DRL to AFC.

\subsection{Deep reinforcement learning and active flow control: challenges and prospects} \label{sec:subsec_DRLinAFC}

Within the field of fluid dynamics, the deep reinforcement learning methods can be applied to a variety of problems. Historically, the first one considered consisted in the behavioral study of swimmers in shoal formations (see Refs.~\onlinecite{gazzola2016,Novati2017,Verma2018} and related works), where the DRL algorithm could choose the position and movements of one/multiple swimmer(s) placed in the wake of a leader. Not only did those studies enhance the scientific knowledge on complex energy saving processes, such as the interception and optimization of vortices by the followers, but it may also have engineering repercussions and benefits in real-world applications~\cite{2010Whittlesey}. Hence, these successful results gave a first overview of the capability and efficiency of DRL algorithms to handle complex physical problems, and paved the way to multiple applications in the field of fluid dynamics~\cite{2020Rabault_reviewDRL_AFC, VinuesaBrunton2021, Brunton2020, 2021LiZhang}.

Nowadays, numerous DRL agents have been implemented in order to, e.g., optimize the position of a small static (e.g., Ref.~\onlinecite{2021Garnier}) or moving (e.g., Refs.~\onlinecite{FanKarnia2020,Xu2020ActiveFC}) body around a larger one with the prospect of reducing the recirculation bubble/the drag downstream, to directly modify the shape of the principal body~\cite{2021Jiao, 2020Viquerat, Yan2019, Yonekura2019} or to control its movements~\cite{Tokarev2020, 2021Ren_hydroStealth} with the same objective of optimizing its aerodynamic/hydrodynamic properties. Other research works handle heat transport issues such as the Rayleigh-Bénard instability~\cite{2020Beintama}, or other convectively-unstable flows~\cite{xu_zhang_2023}.

\citet{2019Rabault-al} led the first study applying a DRL algorithm (PPO) to an AFC problem. Two small synthetic jets were placed symmetrically on the top and bottom of a cylinder immersed in a 2D flow, aimed at shrinking the instabilities behind the body (i.e. the Kármán vortex street). This bi-dimensional showcase, inspired by the original setup of \citet{Schafer1996}, is now a generic benchmark used to test the recent developments of DRL in the field, and has therefore known great and recurrent interest~\cite{2019RabaultKuhnle, 2020Tang_RobustAFC, 2021Ren_AFCturb, 2022Varela}. Other studies also consider AFC problems using actuators, to control the flow separation~\cite{Shimomura2020, batikh:hal-01820331} or to limit the skin friction in wall-bounded channels~\cite{2022Hasegawa, Guastoni2023}, to cite a few.
\\

The community now redoubles its efforts on increasing the complexity of the AFC problems handled with DRL. The first upcoming milestone consists in proving the efficiency of the existing algorithms when it comes to highly turbulent conditions. Issues of instabilities in turbulent boundary layers~\cite{Xu2017_2, Xu2017_1} (TBLs) would for instance become reachable -- then with fruitful engineering prospects (see, e.g., Ref.~\onlinecite{2021Yu_GA_for_TBLs} for the control of a TBL with genetic algorithms). For this purpose, successive research works have for instance increased the Reynolds number in the originally-laminar cylinder problem treated in Ref.~\onlinecite{2019Rabault-al} with a Reynolds number of $Re = 100$. They successfully proved the efficiency of PPO in a range $Re \in [60;400]$~\cite{2020Tang_RobustAFC}, then for the first time in a weakly turbulent condition ($Re = 1000$)~\cite{2021Ren_AFCturb} and recently reached $Re = 2000$~\cite{2022Varela}. In Ref.~\onlinecite{2022Varela}, the algorithm suggests a strategy that differs significantly from the opposition-control solution. PPO is often chosen for its relative simplicity while having comparable performances to the other DRL algorithms, and because flow control issues generally require continuous actuation (thus not reachable with DQN for instance). But DDPG also raised promising results. In Ref.~\onlinecite{2019Bucci}, the authors stabilized a chaotic system governed by the 1D Kuramoto-Sivashinsky equation with only partial observations of the environment, thereby answering some concerns of the community on the efficiency of RL methods in partial observations situations~\cite{Recht2019}. In their appendix, \citet{2020Beintama} also confirmed the efficiency of DRL (PPO) to control chaotic systems (the Lorentz attractor). These two results are therefore encouraging with regard to the increase and control of turbulences by DRL.

A second milestone to reach is the enlargement of the situations/flows~\cite{2021Ren_AFCturb}, going to three dimensional simulations for instance. A few works have already been led in three dimensions (see, e.g., Refs.~\onlinecite{Kuhn2011,Kuhn2011_part2,FanKarnia2020} for model-free methods, or Ref.~\onlinecite{brackston2016} for a model-based application). A pioneering exploitation of invariances by \citet{belus2019} could enable DRL algorithms to deal with an arbitrary large number of actuators while highly limiting the increase of the training cost, and thereby give access to larger AFC problems. More generally, the study of symmetries, equivariances and invariances may enable one to consider larger and more turbulent flow conditions, but it will require a non-trivial implementation. Indeed, if the environment holds symmetries, or if the governing equations contain invariances or equivariances, the ANN will not naturally exploit them. For instance, with a 'naive' ANN, one is not sure that the agent will preserve theoretically-equivariant actions in reality~\cite{ZengGraham2021}. Nonetheless, with sufficient training data, the ANN may learn invariant and equivariant properties by setting some constraints between the weights of the network, but it will then reduce its degrees of freedom~\cite{Ravanbakhsh2017, Sannai2019}. Hence, if one wants a constant number of degrees of freedom and take into account equivariant and invariant properties, one may require a larger ANN, and thus more training data. To prevent such an increase, one has to support the ANN to take advantage of the invariances, equivariances and symmetries when they exist, by encoding a state or state-action reduced space in which the agent can work equivalently to the real environment. \citet{ZengGraham2021} encoded a 'symmetry-reductor' applied to a 1D Kuramoto-Sivashinsky environment controlled by DDPG. They found robust strategies stabilizing the chaotic system in the state-action symmetry-reduced subspace, and proved the enhanced efficiency of DDPG in this subspace in comparison with the 'fully-developed' space. Their work is an encouraging milestone illustrating  how important it will be to consider equivariances, invariances and symmetries when they exist in chaotic/turbulent flow control problems, and how to do so. Eventually, by defining strategies of cooperation between multiple parallel agents, multi-agent reinforcement learning (MARL) has recently proved its efficiency in AFC~\cite{Bae2022, Guastoni2023}, and may help enhancing the complexity of the flows considered.
\\

Encouraging avenues for the raise and enlargement of complexity are thus already under study. In addition, the aforementioned parallelization methods (see section \ref{sec:Parallelization_in_DRL}, and, e.g, Ref.~\onlinecite{2019RabaultKuhnle} for a direct application in AFC) are also promising, as the number of episodes needed for the training of the NN increases in proportion with the complexity of the flow. Both invariances and parallel DRL should be considered to deal with large turbulent flows. Transfer learning~\cite{TaylorStone2009} is also of great interest. Multiple studies have already illustrated the effectiveness of DRL-learned strategies in situations different from the ones considered in the training phase (see for instance Refs.~\onlinecite{2020Tang_RobustAFC,2021Ren_AFCturb,2021Ren_hydroStealth}), and their robustness when confronted to an input or output noise (e.g., Refs.~\onlinecite{xu_zhang_2023,ZengGraham2021}). This ability the agent has to transfer its learning to a wider range of flow conditions proves -- once more -- the robustness of DRL algorithms and their advantageous properties compared to other AFC methods. Moreover, a recent work~\cite{Paris2022_bis} has shown an interest in reducing the number of actuators while limiting the deterioration of the control sequence, exploiting the malleability of NNs. It could thus enable one to deal with larger problems without increasing the number of actuators. Eventually, DRL as well as machine learning in general may also enhance knowledge on physical processes where the classical mechanics predictions encounter limitations~\cite{Noack2019_Book_EnhanceKnowledge}, such as the behavior of chaotic systems. Taking the example of AlphaGo~\cite{Silver2016}, not only did the algorithm impress the community for its performances, but also because it helped understanding underlying structures in the game, for instance by choosing strategies commonly thought as inefficient until then. Therefore, when applied to chaotic systems, DRL methods may highlight unknown structures and discover new physical properties. The non-opposition-control solution found by PPO in Ref.~\onlinecite{2022Varela} goes in this direction. In other words, DRL methods may help enhance our knowledge on physical systems~\cite{RabaultKuhnle2022_book}, meanwhile our current knowledge needs to be exploited to encode better algorithms, taking into account symmetries and invariances for instance.

Hence, data-driven methods, and more particularly DRL algorithms, have already demonstrated unrivalled performances in active flow control problems. They now have to handle larger and more complex situations, as the latter would raise even more engineering applications and scientific knowledge.
\\

As a last comment, we would emphasize the importance of open-sourcing the codes of the DRL algorithms applied to flow control problems. As it has been developed in this review, fruitful methods may emerge from combinations of existing algorithms. For a reproducibility purpose, and because technical implementation choices may have a significant influence on the results, sharing the training data alongside the code and the theoretical details seems to be essential~\cite{RabaultKuhnle2022_book}. For examples of shared toolkits, the reader may refer to Ref.~\onlinecite{Wang2022_OpenFOAM} and to Ref.~\onlinecite{2022Maceda_xMLC}, which respectively present an open-source Python platform for DRL and an open-source GP technique for fluid mechanics.

\section{Concluding remarks} \label{CR}

In the present review, a general overview of the main RL concepts and of the DRL framework has first been given. The potential of DRL for the active control of a flow has then been highlighted -- comparatively to either reactive or predetermined other methods. Not only may the DRL effectiveness be useful in multiple engineering applications, but it could also help science in general. Indeed, this exhaustive exploration tool may provide knowledge in theoretical-physics fields similarly to what it gave to the Go-game-community: the discovery of previously unknown underlying structures in the studied system.

For 200 years, science has focused on the application of analytical methods. This often implies local linearization, as well as stability and modal analyses. However, in many instances this is not meaningful, because there is no such thing as a clearly defined base flow that is similar enough to instantaneous snapshots to perform linearization. Therefore, full-scale  applications may be far enough from a well-defined configuration that linearization is not representative. Consequently, in general, when handling non-linear high-dimensional systems, there is no choice but to study directly the full system. Thus, most analytical tools cannot be used anymore, and this is one of the reason why methods such as DRL are attractive: they can perform very well in much more general conditions than traditional methods, at the cost of being very data intensive. Hence, DRL represents a methodological shift. It will take time before we as a community understand the full value and impact of these methods in AFC, and we are currently in the middle of a massive effort to investigate how far DRL can be taken for AFC applications, and what new insights can be gained from that.

\section*{Acknowledgements}
R.V. acknowledges financial support from ERC grant no. `2021-CoG-101043998, DEEPCONTROL'. Views and opinions expressed are however those of the author(s) only and do not necessarily reflect those of the European Union or the European Research Council. Neither the European Union nor the granting authority can be held responsible for them.


\bibliography{Biblio}

\end{document}